\documentclass[lettersize,journal]{IEEEtran}
\usepackage{algorithm}
\usepackage{algpseudocode}
\usepackage{algorithmicx}
\usepackage{array}
\usepackage[caption=false,font=normalsize,labelfont=sf,textfont=sf]{subfig}
\usepackage{textcomp}
\usepackage{stfloats}
\usepackage{url}
\usepackage{verbatim}
\usepackage{graphicx}
\usepackage{cite}
\graphicspath{{Images/}}
\DeclareGraphicsExtensions{.pdf,.jpeg,.png}
\usepackage{etoolbox}
\usepackage{float}
\usepackage[english]{babel}
\usepackage{adjustbox}
\usepackage{amssymb}
\usepackage{amsmath}
\usepackage{amsthm}
\usepackage{mathtools}
\usepackage{relsize}
\usepackage{ellipsis}
\usepackage{textcomp}
\usepackage{cite}
\usepackage{graphicx}  
\usepackage{setspace}
\usepackage{esint}
\usepackage{booktabs}
\usepackage{subcaption}
\usepackage{graphicx}
\usepackage{setspace}
\usepackage{enumitem}
\usepackage{epsfig}
\usepackage{epstopdf}
\usepackage{caption}
\usepackage{psfrag}
\usepackage{grffile}

\usepackage{color}
\usepackage{hyperref}
\usepackage[english]{babel}
\usepackage{amsthm}
\usepackage{makecell}
\usepackage{tabularx}
\usepackage[table, svgnames, dvipsnames]{xcolor}
\usepackage{makecell, cellspace, caption}
\usepackage{algpseudocode}

\begin{document}{
\title{LoRaConnect: Unlocking HTTP Potential on LoRa Backbones for Remote Areas and Ad-Hoc Networks}
\author{Atonu~Ghosh,~\IEEEmembership{Graduate~Student~Member,~IEEE,}
        Sudip~Misra,~\IEEEmembership{Fellow,~IEEE}

\thanks{Atonu Ghosh and Sudip Misra are with the Department of Computer Science and Engineering, Indian Institute of Technology Kharagpur, Kharagpur 721302, India (e-mail: atonughosh@outlook.com; sudip\_misra@yahoo.com).}

}

\maketitle
{\color{black}
\begin{abstract}
    Minimal infrastructure requirements make LoRa suitable for service delivery in remote areas. Additionally, web applications have become a de-facto standard for modern service delivery. However, Long Range (LoRa) fails to enable HTTP access due to its limited bandwidth, payload size limitations, and high collisions in multi-user setups. We propose \emph{LoRaWeb} to enable HTTP access over LoRa. The LoRaWeb hardware tethers a WiFi hotspot to which client devices connect and access HTTP resources over LoRa backhaul. It implements caching and synchronization mechanisms to address LoRa’s aforementioned limitations. It also implements a message-slicing method in the application layer to overcome LoRa's payload limitations. We evaluate the proposed system using actual hardware in three experimental setups to assess the baseline performance, ideal scenario, and practical application scenario with Frequency Hopping Spread Spectrum (FHSS). Additionally, it implements a ``ping’’ operation to demonstrate Internet capability and extensible nature. LoRaWeb achieves an average throughput of 1.18 KB/S, with an access delay of only $\approx 1.3 S$ for a 1.5KB webpage in the baseline setup. Moreover, it achieves an access delay of $\approx 6.7 S$ for a 10KB webpage in the ideal case and an average end-to-end delay of only $\approx 612$ ms in the FHSS-based setup. Comparison with benchmark suggests multi-fold improvement. 
\end{abstract}


\section{Introduction}

\IEEEPARstart{L}{oRa} has established itself as one of the key enablers of the Internet of Things (IoT) \cite{9854120} due to its low-power and long-range capabilities\cite{huan2023one}. Additionally, LoRa offers a range of up to a few kilometers \cite{wu2023low}, which can be leveraged to connect distant locations or systems. Due to its limited bandwidth, it is suitable for services where small amounts of data are exchanged. Additionally, the minimal infrastructure requirements of LoRa make it an excellent choice for emergency and ad-hoc networks. 

Web applications have become a de-facto standard for service delivery. The delivery of web applications over LoRa holds tremendous potential in almost all aspects of modern human life. With this amalgamation, far-flung locations can be connected to the mainstream networks and help alleviate the sense of deprivation among the citizens. Existing Low-Power Wireless Area Network (LPWAN) technologies, such as the Low-Power Long-Range Wide Area Network (LoRaWAN), require additional infrastructure and heavily suffer due to overheads. It restricts fine-grained control for enhancing payload size, caching mechanisms, and synchronization. However, rural and distant locations severely lack the basic infrastructure, especially in developing and under-developed nations \cite{bhuiyan2022design}. As a result, these locations remain disconnected and are often deprived of basic and essential services.  

LoRa can help address this by enabling the delivery of lightweight web applications in a decentralized manner with no additional infrastructure requirements. However, LoRa natively does not support the HTTP protocol as it necessitates high data rates and limitations on the payload size. Furthermore, due to the narrow bandwidth, LoRa suffers from prohibitively high collisions in multi-client setups. 

\textbf{Example Scenario:} This work considers isolated rural areas that lack critical modern infrastructure for communication. Unlike the urban locations, these areas often miss the essential digital media-based bulletins published by the local Government and non-Government organizations due to the poor communication facilities. Fig. \ref{fig: system_overview} depicts the example scenario. LoRaWeb enables the authorities to upload the announcements in the LoRaWeb server deployed in the village office. The residents carry the LoRaWeb device and use smartphones to access the bulletin web pages over the intra-village LoRa network. The LoRa network acts as a backbone and exchanges data to and from the sender and receiver LoRaWeb devices.

\subsection{Motivation} \label{motivation}
Lately, LoRa has witnessed several research efforts that enhance the range and energy consumption. Several solutions have been proposed to leverage the low-power and long-range capabilities in smart cities, smart factories, etc. However, these studies do not address the challenges in multi-user setups with high collisions, larger payloads, and HTTP access. The work in \cite{10.1007/s11036-019-01235-5} is closely related to our work, and it implemented a messaging system over LoRa. It uses a Raspberry Pi gateway to connect to the Internet and integrate the Telegram platform with its messaging system. However, the work fails to deliver complete web applications over LoRa, and the study fails to evaluate multi-user scenarios over a single LoRa channel. Our work ``LoRaWeb" addresses the lacunae in the existing LoRa-based solutions.  
}

\begin{figure}[h]
\centering
\includegraphics[width=0.6\columnwidth]{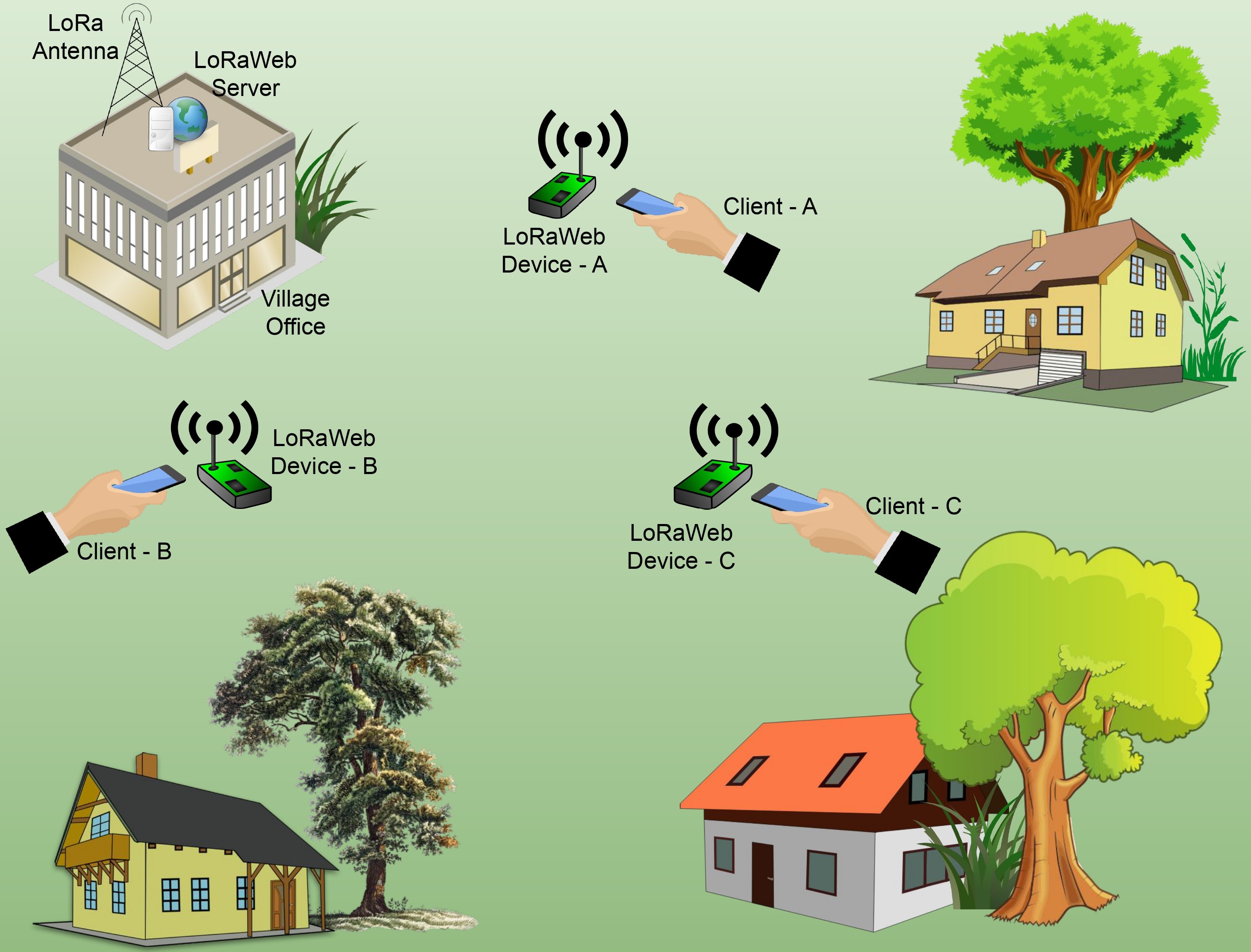}
\caption{Overview of the proposed system deployed in a village with LoRaWeb node configured as server and client nodes}
\label{fig: system_overview}
\end{figure}

\subsection{Contributions}
The specific contributions of this work are as follows:
\begin{enumerate}
    \item This work implements an end-to-end system to deploy HTTP protocol over a LoRa network {\color{black}to deliver complete web applications}. Furthermore, it furnishes the detailed results obtained from an actual hardware-based {\color{black}multi-client} deployment of the proposed mechanisms.

    \item {\color{black}Also, it implements an asynchronous and FHSS-based approach to handle multiple clients and requests over a single channel LoRa link.}

    \item {\color{black}It} proposes and implements a LoRa application layer-based message-slicing methodology to overcome the hardware limitations on message length.

    \item {\color{black}It also implements a ``ping" mechanism on the proposed system to demonstrate the possibility of the Internet and the extensibility of the system.}
    
    \item Furthermore, this work demonstrates the methods to route requests received on WiFi and map them to the corresponding resources on the LoRa network. 

    \item Finally, the work proposes and implements a content caching {\color{black}and synchronization} mechanism to reduce the number of transmissions in the LoRa network.

\end{enumerate}

\section{Related Work} \label{Related Work}
Several researchers have developed IoT solutions using the LoRa technology for {\color{black}smart agriculture \cite{s23052725}}, air quality monitoring \cite{7496973}, fire detection \cite{9121967},  slope monitoring \cite{9635758}, smart water and energy metering \cite{10196447}, and industrial machinery monitoring \cite{10015016}. 

\subsection{Improving Efficiency of LoRa Networks}
As LoRa is susceptible to collisions, several works have proposed methods to address this. Researchers in \cite{10.1145/3452296.3472931} proposed the ``Concurrent Interference Cancellation (CIC)" technique, which, unlike traditional methods, introduces a symbol-by-symbol decoding approach. It leveraged sub-symbol variations to cancel interfering frequencies and operated parallelly, increasing the throughput. CIC achieved ten times improvement in an actual hardware-based deployment. To address inter-network interference and to enhance LoRaWAN's performance in concurrent transmissions, researchers in \cite{9155433} proposed ``Online Concurrent Transmissions (OCT)". OCT improved the throughput by recovering the packets that collided at the gateway. For this, OCT used preamble detection, symbol recovery, and cross-decoding based on time and power offsets. The researchers achieved a data reception rate of over $90\%$ through an actual hardware implementation. For real-time collision detection with low SNR signals, another group of researchers proposed ``Pyramid" \cite{9488695}. In this, they tracked the peaks across moving windows and extracted frequency-domain features that are resistant to noise. Pyramid achieved accurate packet separation even under severe collisions by grouping symbols, tracking peak trajectories, and estimating their apex.

\subsection{Cross Technology Communication (CTC)} }
{\color{black}{Interfacing LoRa with other technologies has become prominent in building more intelligent and scalable systems. Towards this, to facilitate the delivery of time-critical messages between LoRa and IEEE 802.15.4 devices in industrial setups, researchers proposed a slotframe structure that enabled message delivery without handshaking with the neighboring nodes in the network. They developed a transmission scheduling algorithm that ran on each device in the network \cite{10228891}. Another group of researchers proposed solutions that integrate LoRaWAN into 5G networks where the Radio Access Network (RAN) is the LoRaWAN RAN, and the backbone network is based on 5G. Additionally, they proposed authentication schemes for primary and secondary authentication purposes \cite{10103169}. Another group of researchers in \cite{wang2023ligbee} proposed ``LigBEE" a novel technique for CTC from LoRa to ZigBee at the symbol level without requiring multiple radios. It leveraged LoRa's chirp symbol's phase pattern, which enabled direct decoding by ZigBee’s OQPSK demodulator. An actual hardware-based implementation demonstrated ``LigBEE"'s effectiveness. Researchers also proposed ``WiRa" \cite{9796831}, which emulates LoRa chirps using WiFi and achieved a throughput of 40.037 kbps.}}\\

{\color{black}{\subsection{Artificial Intelligence(AI)/Machine Learning(ML) for LoRa}
Intelligence(AI)/Machine Learning(ML) for LoRa}
As cache access time impacts system performance and throughput, researchers in \cite{9060811} proposed a framework that implemented distributed learning in edge devices to process data and make decisions collaboratively. This recommendation system for edge systems stored correlated contents to improve storage utilization and enhance access latency. Researchers in \cite{10050828} proposed ``AI-ERA", which used a Deep Neural Network (DNN) model trained on simulation data to optimize resource allocation. AI-ERA predicted the optimal parameters for data transmission, using which it achieved a packet success ratio of $32\%$ in static environments. To further enhance the energy efficiency of LoRa networks, researchers in \cite{9802526} proposed a deep learning-based double-training method. They dynamically adjusted the data transmission power in LoRa networks. For this, the researchers first trained the artificial neural network (ANN) using data derived from simple models and then using datasets generated via Monte Carlo simulations. As a result, they achieved a ten times reduction in the mean square error.}

{\color{black}{\subsection{Synthesis}
Although there have been significant advancements in wireless networks and LPWANs as witnessed in the literature, we found several lacunae. A large number of the works in the literature are simulation-based, and they fail to address the requirements in remote and disconnected regions that lack critical infrastructure. There is a lack of systems and methods for enabling the access of commonly used protocols, such as HTTP, over long-range and low-power networks such as LoRa. Additionally, the related works do not provide details of content caching in resource-constrained LoRa networks to enhance access times.}}


\section{LoRaWeb Node} \label{loraweb_node}
LoRa nodes are the primary constituent elements of the network for web service access. A network ($\mathbb{N}$) for web service access is a tuple consisting of a web server ($\mathbb{S}$) and $k$ number of client devices ($\mathbb{D}$). It is denoted as 

\begin{equation}
  \mathbb{N(S,D)} =
    \begin{cases}
      \mathbb{S} & \geq 1, \in \mathbb{I}^+\\
      \mathbb{D} & = \left\lbrace \mathbb{D}_1, \mathbb{D}_2, \dots, \mathbb{D}_k\right\rbrace k \geq 1, \in \mathbb{I}^+\\
    \end{cases}       
\end{equation}

{\color{black}{This work utilizes the LoRa node developed in \cite{10142009} to build the network. The node comprises a low-power microcontroller that features an onboard WiFi chip. The microcontroller integrates with a LoRa module and an antenna to enable LoRa communication capabilities on the same hardware. Section \ref{experimental_setup} provides a detailed description of the experimental network setup used in this work. The LoRa node has two modes of operation, i.e., as a web server and a client device.}}

\subsection{As LoRaWeb Client Device} \label{client_device}
\begin{figure}
    \captionsetup{labelfont={color=black}}
    \centering
    \includegraphics[width=0.45\columnwidth]{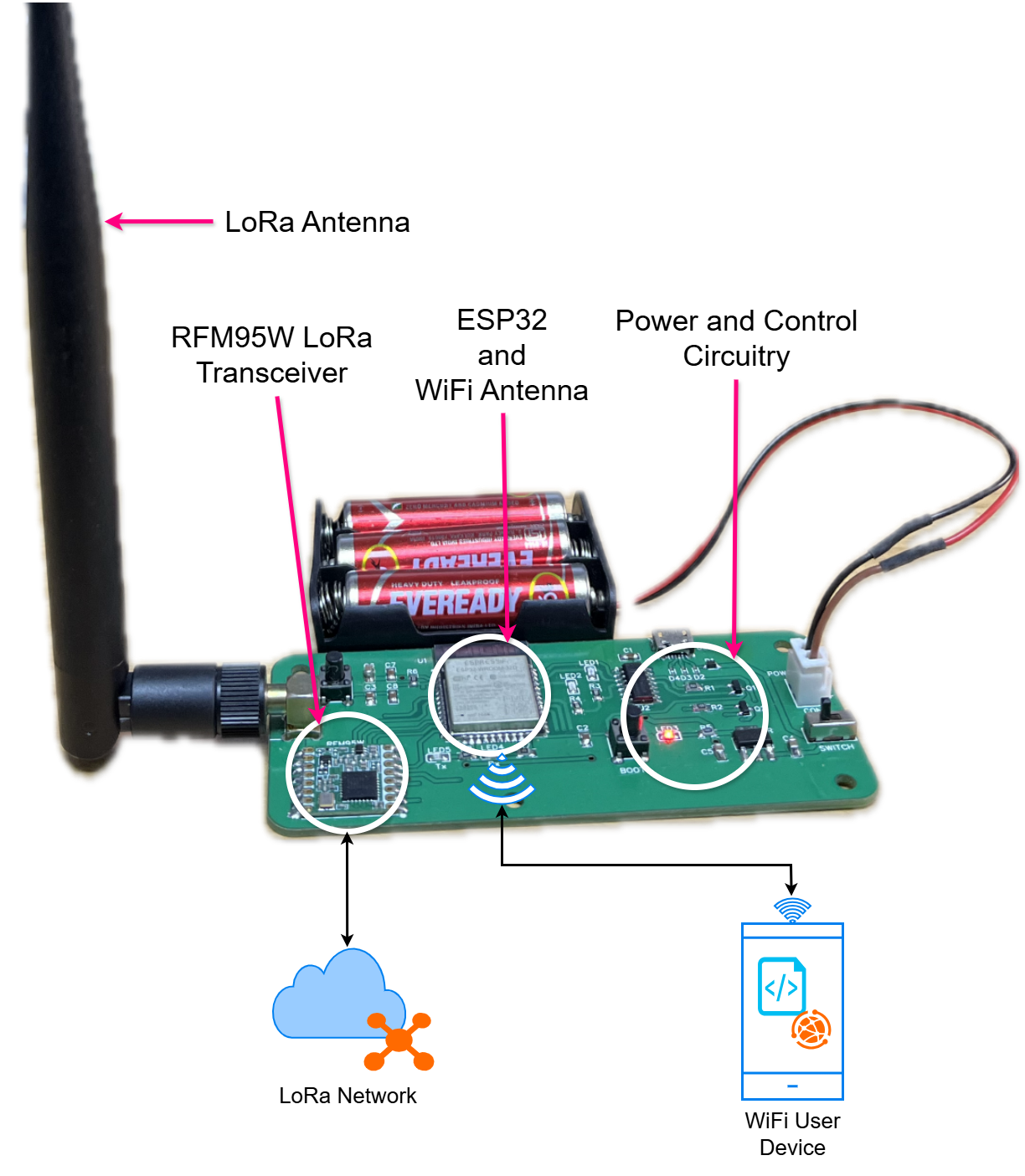}
    \caption{{\color{black}{Composition of LoRaWeb node and interaction with WiFi device and LoRa network }}}
    \label{fig: home_node}
\end{figure}

When configured as a client device, a LoRaWeb node bridges the LoRa and WiFi communication channels in the hardware. The onboard WiFi chip in the LoRaWeb hardware tethers a WiFi access point to which users connect devices such as smartphones to access the network through a web browser. The mobile client devices provide easy access to the LoRa network through commodity user devices. The LoRaWeb device routes the requests received on the WiFi channel to access the corresponding resources on the LoRa network server. Fig. \ref{fig: home_node} depicts the interaction among the user devices with the LoRa network through the WiFi communication channel.  

\subsection{As LoRaWeb Server} \label{loraweb_server}
As a web server, the LoRa node stores the web pages in its memory when uploaded or updated by a user. Upon receiving requests from client devices over the LoRa link, it responds with corresponding content. It executes the proposed receiver-transmitter synchronization and message-slicing algorithms while transferring data to the client devices. Unlike the LoRaWeb client device, the server configuration does not employ the WiFi chip in the hardware. It only communicates over the LoRa link to receive and respond to requests.


\section{LoRaWeb Access Methods} \label{message_routing}

\subsection{Request Handling} \label{request_handling}
The LoRa client node accepts Uniform Resource Locator (URL) addresses over the WiFi link to fetch web pages from the LoRa backbone network. Hence, to enable the user to enter URLs, the client node implements a TCP/IP socket program that listens for addresses on port number 80. As the user enters a URL to request a resource in the network, it creates a request object. The request object contains information such as the request method, requested URL, host address, connection type, user agent software, accepted resource type, encoding type, and accepted language.

\begin{figure} [h]
    \centering
    \includegraphics[width=0.45\columnwidth]{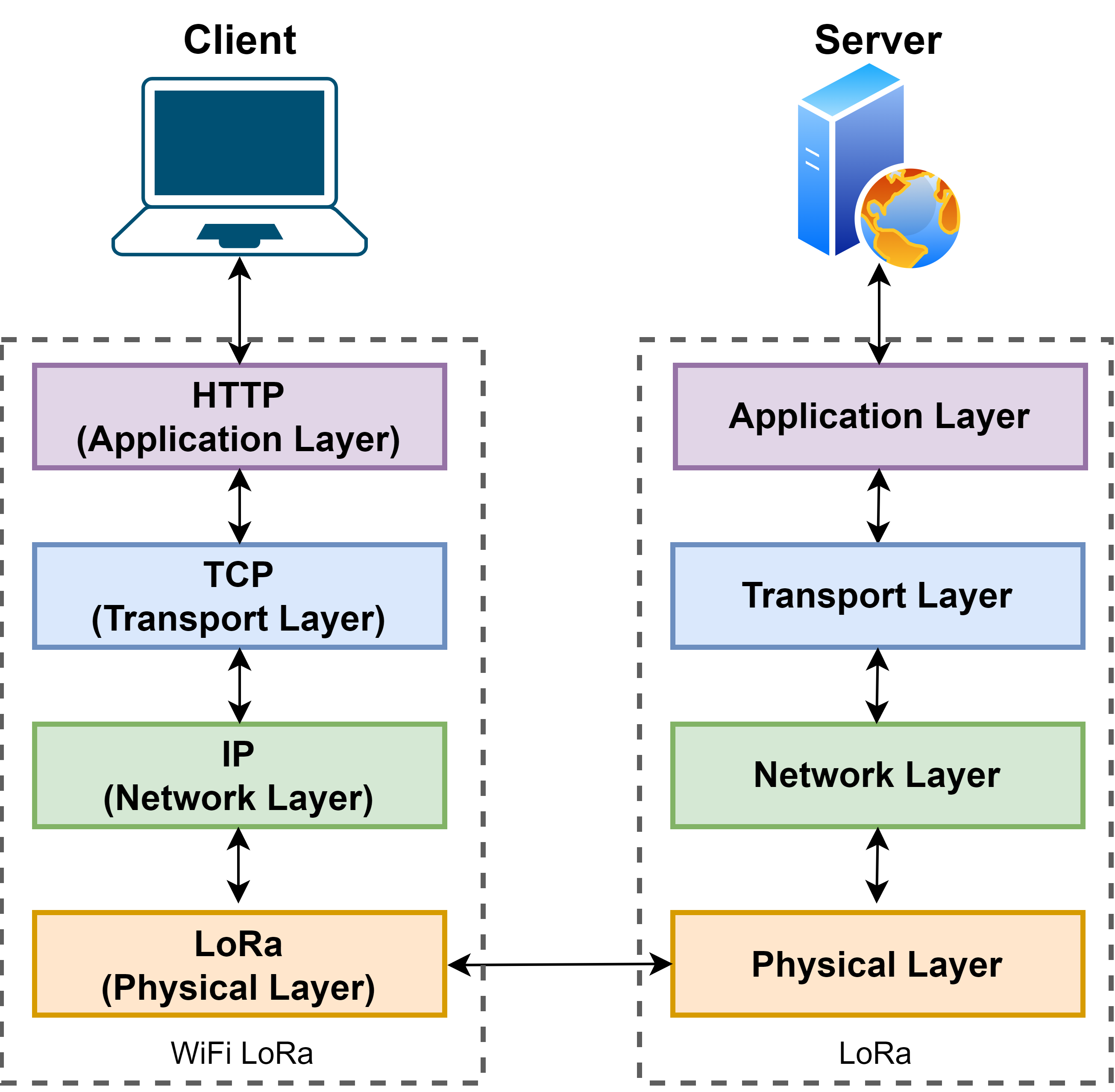}
    \caption{Layered architecture of the request-response flow between client and server in the LoRaWeb network}
    \label{fig: request_forwarding}
\end{figure}

 The client node parses the contents of the request object, filters the URL, and forwards it over the LoRa link to reach the server for resource mapping and further processing. Fig. \ref{fig: request_forwarding} depicts the layered architecture and the interoperation of the client and server hardware exchanging the URL requests and responses in such a configuration.\\

\newtheoremstyle{customdefinition}%
  {3pt}{3pt}{}{}{\bfseries}{.}{ }{} 

\theoremstyle{customdefinition}
\newtheorem{definition}{Definition}
{\color{black}
\begin{definition}[\textit{Response Time}]
It is the total time a LoRaWeb client node takes to send a request to the LoRaWeb server and receive the first packet/chunk in response.
\end{definition}
}

{\color{black}\subsection{Multiple Concurrent Request Handling} \label{concurrent_request_handling}
User devices such as smartphones connect to the LoRaWeb client over the WiFi hotspot tethered by it and send HTTP requests. The LoRaWeb client comprises an asynchronous web server that receives concurrent HTTP requests for the same or different resources. The LoRaWeb client node serves identical requests immediately by returning the resources from its cache memory. It also maintains a user-configurable and optional timeout to maintain the freshness of the data. On a timeout, it compares the versions on the LoRaWeb server and in the cache. This process drastically reduces constrained LoRa channel access and improves the response time. However, the LoRaWeb client device queues the distinct web page requests and resolves the requests one at a time over the LoRa backhaul.

\subsection{Receiver Transmitter Synchronization} 
\label{rx_tx_sync}
{\color{black}
Sender-receiver synchronization is essential to minimize message loss. A client node in the LoRaWeb system initiates a request to access resources on the server. As described in Algorithm \ref{synchronize_nodes}, the server node looks up for the requested resource, i.e., the webpage, and slices the payload. During this time, the client waits for messages from the server to start receiving the chunks. The server awaits an acknowledgment (ACK) after transmitting each chunk. If the ACK does not arrive within a user-configurable duration, it resends the last chunk as it assumes message loss. If a subsequent ACK is lost, the server dynamically adjusts the timeout duration and continues. The number of retries is also user-configurable and allows to avoid excessive retransmissions. The receiver resends an ACK with the ID of the last received chunk in case an expected chunk is not received within a timeout. Moreover, the receivers do not listen to incoming packets all the time. It only listens for transmissions until the end keyword \emph{$<$/html$>$} is received. The receiver stops listening after a timeout when the keyword is not received due to synchronization faults. As the number of transmissions depends on the number of pieces of the message to be transferred, which in turn depends on the payload size, the algorithm executes in $O(n)$ time.}

\begin{algorithm} 
\caption{\textcolor{black}{Synchronize and Receive}}
\label{synchronize_nodes}
\textcolor{black}{
\begin{algorithmic}[1]
    \renewcommand{\algorithmicrequire}{\textbf{Inputs:}}
    \renewcommand{\algorithmicensure}{\textbf{Output:}}
    \Require{$S$ (sender node), $R$ (receiver node), $U$ (URL)}
    \Ensure{$O$ (status of message transfer)}
    \Procedure{Sync\_And\_Transmit}{$S, R, U$}
        \State \textbf{Send\_Request}($S, R, U$)
        \While{$retries \leq MAX\_RETRY$ \&\&\ $!TxDone$}
            \State $chunk \gets RECEIVE\_CHUNK()$
            \If{$chunk.id > lastAckedID$}
                \State \textbf{Process Chunk}($chunk$)
                \State \textbf{Send\_ACK}($chunk.id$)
                \If{$chunk$ ENDS WITH ``$<$/html$>$''}
                    \State $TxDone \gets True$
                \EndIf
            \ElsIf{\textbf{Timeout Detected}}
                \State \textbf{Resend\_ACK}($lastAckedID$)
                \State $retries++$
            \EndIf
        \EndWhile
        \If{$TxDone$}
            \State $O \gets ``0"$
        \Else
            \State $O \gets ``1"$
        \EndIf
        \State \textbf{return} $O$
    \EndProcedure
\end{algorithmic}
}
\end{algorithm}
}

\subsection{Caching Web Pages}
\label{caching_web_pages}
The delay in fetching the contents from a server using a URL is directly proportional to the number of transmissions required to fulfill the request. Hence, to minimize this delay, the LoRaWeb system optionally stores the previously accessed web pages in its memory.

It maintains the cached web page names in a dictionary ($\mathbb{L}$) as ordered pair of keys ($\mathbb{Q}$) and values ($\mathbb{R}$) as presented in Eq. \ref{eq:2} in the local memory to facilitate the lookup of a web page in the cache. The keys are the web page names, and the values are the version numbers. The client device checks this dictionary for earlier access for each request. If the currently requested URL is present in the dictionary of accessed URLs, then it checks the web page version number from the server. 

\begin{equation}
 \label{eq:2}
  \mathbb{L} =
      (q_i,r_j)      
\end{equation}

where, $q \subseteq \mathbb{Q}, r \subseteq \mathbb{R}$, $|Q| \geq 1, \in \mathbb{I}^+$, $|R| \geq 1, \in \mathbb{I}^+$, and $i,j \geq 1, \in \mathbb{I}^+$.  

It compares the version number retrieved from the local dictionary with the version number fetched from the server. If the version numbers match, it fulfills the request by simply returning the web page's contents from its local memory. Otherwise, it invokes Algorithm \ref{synchronize_nodes} to fetch the updated version of the web page from the server. Algorithm \ref{cache_webpages} describes the caching mechanism implemented in the LoRaWeb system. {\color{black}This ensures that the latest content is delivered for each request. Thus maintaining the freshness of data. Also, the LoRa channel is completely avoided in multiple requests for the same resource which results in a reduction of contention of the LoRa channel. This also greatly reduces the LoRaWeb system's response time.} LoRaWeb implements the dictionary using a hash table which results in a lookup time of only $O(1)$ {\color{black}which is highly optimized for resource-constrained microcontrollers.}

\begin{algorithm}
\caption{Cache Web Pages}
\label{cache_webpages}
\begin{algorithmic}[1]
    \renewcommand{\algorithmicrequire}{\textbf{Inputs:}}
    \renewcommand{\algorithmicensure}{\textbf{Output:}}
    \Require{$S$ (sender node), $R$ (receiver node), $U$ (URL), $I$
    \hspace*{1.5em} (dictionary of URLs in the cache)} 
    \Ensure{$W$ (requested web page)}

    \Procedure{Get\_Web\_Page}{$S$, $R$, $U$, $I$}
        \If{$I.IS\_IN(U)$}
            \State{$version_{c} \gets FETCH\_KEY($U$)$ from $I$}
            \State{$version_{s} \gets GET\_VERSION(U)$ from $S$}
            \If{$version_{s} == version_{c}$}
                \State{$W \gets FETCH\_WEB\_PAGE(U)$}
                \State{\textbf{return} $W$}
            \Else
                \State{$W \gets$ $SYNC\_AND\_TRANSMIT$}($S, R, U$)
                \State{\textbf{return} $W$}
            \EndIf
        \Else
            \State{$W \gets$ $SYNC\_AND\_TRANSMIT$}($S, R, U$)
            \State{\textbf{return} $W$}
        \EndIf
    \EndProcedure
\end{algorithmic}
\end{algorithm}

\subsection{Handling Large Payloads}
The hardware buffer size limits the maximum payload size that can be exchanged by the LoRa nodes. Hence, transmitting an arbitrary payload of size greater than the hardware buffer size necessitates efficient handling. This work employs a message payload chunking mechanism that slices the payload into pieces of {\color{black}$250$ bytes each}. 

{\color{black}\subsection{Constant Delays}
The time taken for slicing a web page of a given size and the time to cache the same web page are constant. Moreover, they appear negligible for higher spreading factors such as 12. However, as the transmission times drastically reduce for lower spreading factors such as 7, these constant delays become prominent and may limit the enhancement achieved in end-to-end latency of the system.}

{\color{black}\subsection{Demonstrating Internet Interaction}
Leveraging its extensible nature and paving the way for future research to enable complex web services over LoRaWeb, it implements a lightweight ping functionality.  A LoRaWeb device sends a ping request to another LoRaWeb device connected to the Internet. The second device performs the ping operation and sends back the results over the LoRa network to the requesting device. This proves LoRaWeb's ability to support basic Internet even with its constraints on bandwidth and latency.} \\

\newtheorem{prop}{Proposition}
\begin{prop}
LoRaWeb eliminates packet loss and ensures Quality of Service (QoS).
\end{prop}

\begin{proof}
    The proposed client-server synchronization and transmission algorithm manages the data exchange between the LoRaWeb devices. The re-transmission and acknowledgment mechanisms in the algorithm as mentioned in Section \ref{rx_tx_sync} ensure the delivery of all the slices of a message. The algorithm's number of retries is user-defined and modifiable as per the deployment's QoS requirements. Moreover, the caching mechanism (Section \ref{caching_web_pages}) greatly reduces the access time as compared to direct access from the server.
\end{proof}

\begin{prop}
Irrespective of payload size, LoRaWeb delivers the web page in its entirety.
\end{prop}

\begin{proof}
    The LoRaWeb system overcomes the hardware limitations on the payload length by implementing a message chunking mechanism in the application layer. The proposed mechanism runs efficiently on the resource-constrained microcontroller and executes in considerably low time. Thus, an arbitrary payload is orderly delivered to the requester.
\end{proof}


\section{Experimental Setup} \label{experimental_setup}

{\color{black}
We evaluated the performance of the LoRaWeb system, by conducting experiments with three setups.
\begin{itemize}
    \item First, multiple LoRaWeb clients with a LoRaWeb server.
    \item Second, a single WiFi user device with a LoRaWeb client. 
    \item Third, multiple LoRaWeb clients with a LoRaWeb server using frequency hopping.
\end{itemize}

With the help of these setups, we evaluated the LoRaWeb system in terms of packet drop ratio (PDR), throughput, data rate, response time, access delay, environmental impact, scalability and collision Impact, fairness index, cache read and write time, cache index read and write time, current consumption, and round trip time for ping implementation as detailed in Section \ref{performance_evaluation}.

\begin{definition}[\textit{Access Delay}]
It is the total time a LoRaWeb client takes to send a request to the LoRaWeb server and receive a complete web page in response.
\end{definition}

To evaluate the baseline performance of the proposed LoRaWeb system, in the first setup, four LoRaWeb client nodes were set to concurrently request a $1.5KB$ webpage from the LoRaWeb server. Each node sent requests at random intervals between $8$ and $25$ seconds to mimic real-world load conditions. With this setup, one of the clients and the LoRaWeb server were monitored to record the performance metrics. This setup was experimented with different configurations of SF, BW, and duty cycle. $SF = 7$ with $BW = 250KH_z$ and $500KH_z$, $SF = 9$ with $BW = 250KH_z$, and $SF = 12$ with $BW = 250KH_z$ were tested for $10\%$, $30\%$, $50\%$, and $100\%$ duty cycles. The SNR and RSSI values were also recorded to analyze fading and SF interference in this setup. The efficiency of the caching mechanism was also evaluated by varying the web page sizes. Section \ref{muli_loraweb_client} provides the detailed evaluation.

The second setup mimicked an ideal case where a user device such as a smartphone connected to the WiFi hotspot tethered by a LoRaWeb client node. The LoRaWeb client connected to the LoRaWeb server over the LoRa backhaul network to serve the user requests. We tested this setup by varying the web page sizes between $1.5 KB$ and $10 KB$. Also, in this experiment, the spreading factor and the bandwidth were set to $7$ and $500KH_z$, respectively. This choice of bandwidth and spreading factor was to achieve minimum latency and maximum data rate. The details of the results and analysis of this setup are provided in Section \ref{single_wifi_client_setup}.

The third setup consisted of four LoRaWeb clients connected to a LoRaWeb server over the LoRa backhaul network. The client nodes requested web pages of size $1.5KB$ from the server at random intervals between 8 and 25 seconds. Unlike the other two setups, this setup implemented an FHSS mechanism to minimize the effects of network congestion and interference, as observed in the static frequency setup. We provide a detailed evaluation of this setup in Section \ref{fhss}.
}

\begin{figure}[h]
\centering
\includegraphics[width=0.45\columnwidth]{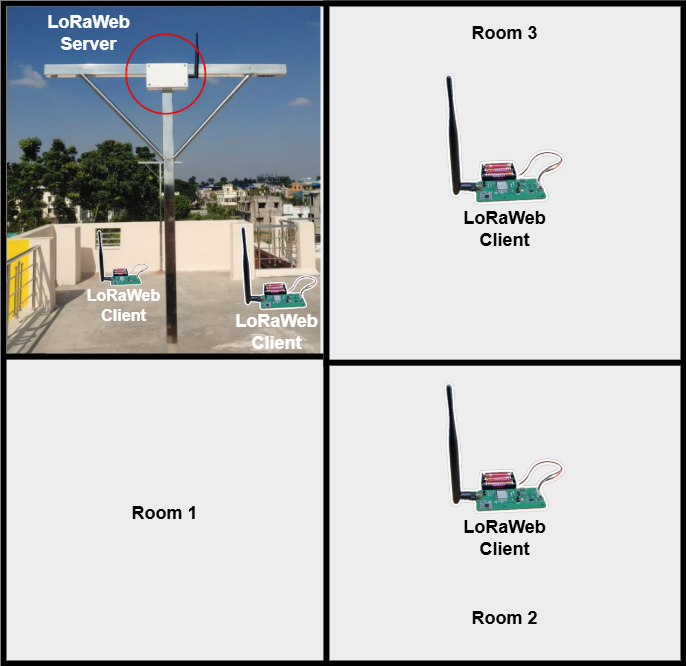}
\caption{LoRaWeb server and client node test deployment}
\label{fig: system implementation}
\end{figure}

The LoRaWeb node illustrated in Fig. \ref{fig: home_node} comprised an $ESP32$ microcontroller embedded on a custom Printed Circuit Board (PCB), an $RFM95W$ LoRa communication module, a LoRa antenna operating at $868MH_z$, {\color{black}and a battery module.}

{\color{black}Figure \ref{fig: system implementation} illustrates the test deployment of the LoRaWeb system. User devices such as smartphones and laptops connected to the LoRaWeb network over WiFi hotspot tethered by the LoRaWeb client. The firmware for both the LoRaWeb client and server was developed in C++. An existing Android web browser was used to test the HTTP access on the LoRaWeb network. Table \ref{experimental_setup_table} lists the detailed configurations of our experimental setup.}

\begin{table}[h]
\centering
\caption{LoRaWeb test deployment configurations}
\label{experimental_setup_table}
\resizebox{0.5\columnwidth}{!}{
\begin{tabular}{| c | c | c |}
\hline \textbf{Item} & \textbf{Parameter} & \textbf{Value} \\
\hline 1 & Network Configuration & Point-to-point \\
\hline 2 & Client Device & \textcolor{black}{Smartphone and Laptop} \\
\hline 3 & Access Protocol & HTTP \\
\hline 4 & WiFi Frequency & $2.4GH_z$ \\
\hline 5 & No. of clients & $1$ and $4$ \\
\hline 6 & Processing Unit (MCU) & ESP32 \\
\hline 7 & LoRa Transceiver & $RFM95W$ \\
\hline 8 & LoRaWeb Firmware & \textcolor{black}{C++} \\
\hline 9 & LoRa Radio & $868MH_z$ \\
\hline 10 & Transmit Power (LoRa) & \textcolor{black}{$17dBm$} \\
\hline 11 & Spreading Factor (LoRa) & \textcolor{black}{$7, 9,$ and $12$} \\
\hline 12 & Bandwidth (LoRa) & \textcolor{black}{$250KH_z$ and $500kH_z$} \\
\hline 13 & Coding Rate (LoRa) & \textcolor{black}{$4/5$} \\
\hline 14 & Test Distance & \textcolor{black}{$5-10 m$} \\
\hline
\end{tabular}
}
\end{table}


{\color{black}

\section{Performance Evaluation} \label{performance_evaluation}

\subsection{Setup-I: Multiple LoRaWeb Clients With A LoRaWeb Server}\label{muli_loraweb_client}

In this setup, we did not include a retry mechanism, which led to some packet losses. This behavior was as per our expectations and it helped all LoRaWeb clients to have fair access to the shared communication channel. With a sufficient number of re-tries, all packets would be delivered as demonstrated in Section \ref{single_wifi_client_setup}, \ref{fhss} and in work by \textit{Magrin et al.} \cite{9294099} where they implemented 20 gateways to achieve a similar success rate of above 90\%. However, this approach leads to the starvation of other nodes due to prolonged channel occupation. We turned-off the caching mechanism to analyze LoRaWeb’s baseline performance. It is to the user's discretion and the deployment's Quality of Service (QoS) requirements to decide whether to implement a re-try and caching mechanisms.

\subsubsection{Packet Delivery Ratio (PDR)} \label{pdr_multi}
For each SF, BW, and duty cycle (10\%, 30\%, 50\%, and 100\%) combination, we recorded the number of chunks successfully delivered ($\alpha$) and calculated the Packet Delivery Ratio ($\lambda$) using Eq. \ref{eqn: pdr_setup_1} where $\beta$ is the total number of packets sent. Table~\ref{tab:pdr_math} provides the calculated PDRs. For 50 successful requests ($\alpha = 50$), the LoRaWeb system sent a varying total number of requests ($\beta$), which depended on the SF, BW, and duty cycle settings.  

\begin{equation} \label{eqn: pdr_setup_1}
    \lambda = \frac{\alpha}{\beta} \times 100
\end{equation}

As the SF increased, the PDR decreased consistently for all configurations. This is due to high Time on Air (ToA) at higher SFs, which resulted in higher channel occupation and contention among the LoRaWeb nodes. Increasing the bandwidth resulted in better PDR, especially for lower SFs. For higher SFs such as $SF = 12$, the effect of bandwidth did not result in appreciable results as the ToA increased. Furthermore, increasing the duty cycle resulted in higher PDRs as observed with $SF = 7$ and $BW = 250KH_z$, the PDR increased from $\approx 53.2\%$ (10\% duty cycle) to $\approx 59.5\%$ (100\% duty cycle). Throughout the experiment, $SF = 7$ and $BW = 500KH_z$, at 50\% or 100\% duty cycle was observed to achieve the highest PDR of $\approx 71.4\%$. 

\begin{table}
\centering
\caption{Packet Delivery Ratio ($\lambda$) for Different Configurations}
\label{tab:pdr_math}
{\color{black}
\resizebox{0.5\columnwidth}{!}{%
\begin{tabular}{|c|c|c|c|}
\hline
\rowcolor[HTML]{EFEFEF} 
\textbf{SF, BW (kHz)} & \textbf{Duty Cycle (\%)} & \textbf{Requests ($\alpha / \beta$)} & \textbf{$\lambda$ (\%)} \\ \hline
7, 250                & 10                      & 50 / 94                             & 53.2                   \\ \cline{2-4} 
                      & 30                      & 50 / 91                             & 54.9                   \\ \cline{2-4} 
                      & 50                      & 50 / 89                             & 56.2                   \\ \cline{2-4} 
                      & 100                     & 50 / 84                             & 59.5                   \\ \hline
7, 500                & 10                      & 50 / 71                             & 70.4                   \\ \cline{2-4} 
                      & 30                      & 50 / 72                             & 69.4                   \\ \cline{2-4} 
                      & 50                      & 50 / 70                             & 71.4                   \\ \cline{2-4} 
                      & 100                     & 50 / 70                             & 71.4                   \\ \hline
9, 250                & 10                      & 50 / 131                            & 38.2                   \\ \cline{2-4} 
                      & 30                      & 50 / 125                            & 40.0                   \\ \cline{2-4} 
                      & 50                      & 50 / 110                            & 45.5                   \\ \cline{2-4} 
                      & 100                     & 50 / 96                             & 52.1                   \\ \hline
12, 250               & 10                      & 50 / 265                            & 18.9                   \\ \cline{2-4} 
                      & 30                      & 50 / 216                            & 23.1                   \\ \cline{2-4} 
                      & 50                      & 50 / 168                            & 29.8                   \\ \cline{2-4} 
                      & 100                     & 50 / 163                            & 30.7                   \\ \hline
\end{tabular}%
}
}
\end{table}

\subsubsection{Throughput}

The throughput values achieved by the LoRaWeb system with different combinations of SF and BW at a 100\% duty cycle is depicted in Fig. \ref{fig: throughput}. It achieved an average throughput of $1.18KB/S$ with $SF = 7$ and $BW = 500KH_z$, which dropped to $\approx 0.78KB/S$ when the bandwidth was set to $250KH_z$. This reduction shows the impact of lower bandwidth on the throughput. We also observed the impact of higher SFs on the throughput as with $SF = 12$ and $BW = 500KH_z$, the average throughput dropped to $0.09KB/S$. Compared to our work, \textit{Kaur et al.} \cite{KAUR2024102844} achieved a maximum throughput of only $\approx 0.44Kb/S$ across spreading factors.

\begin{figure}[h]
\captionsetup{labelfont={color=black}}
\centering
\includegraphics[width=0.7\columnwidth]{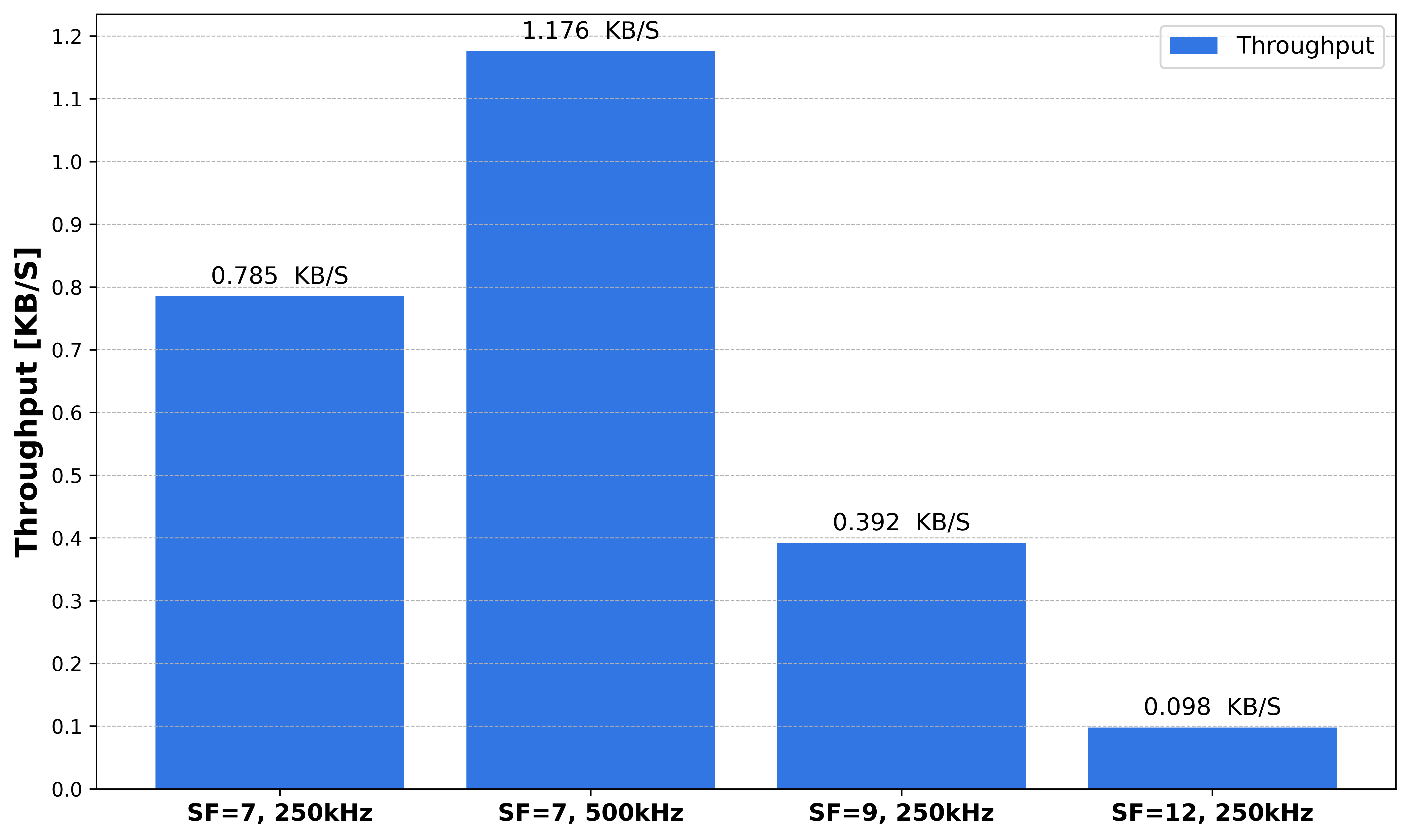}
\caption{{\color{black}Throughput achieved by LoRaWeb for various configurations of SF and BW at 100\% duty cycle}}
\label{fig: throughput}
\end{figure}

\subsubsection{Data Rate}
We experimented with the LoRaWeb system to evaluated the data rates achieved by it which is key parameter indicating its performance levels. Fig \ref{fig: data_rate} depicts the data rates achieved by the LoraWeb system for different combinations of spreading factor and bandwidth at 100\% duty cycle setting. Duty cycles showed minimal impact on the data rates. But, similar to the throughput, the data rate was also affected by spreading factors and bandwidth settings as a result of increased ToA for higher spreading factors. We observed that the data rate dropped from 1.26KB/S to 0.06KB/S when the spreading factor was increased from 7 ($500KH_z$) to 12 ($250KH_z$). We also observed the effect of bandwidth on the data rates. For instance, with a spreading factor of 7, when the bandwidth was increased from $250 KH_z$ to $500 KH_z$, the data rate increased from 0.84KB/S to 1.26KB/S. 

\begin{figure}[h]
\captionsetup{labelfont={color=black}}
\centering
\includegraphics[width=0.7\columnwidth]{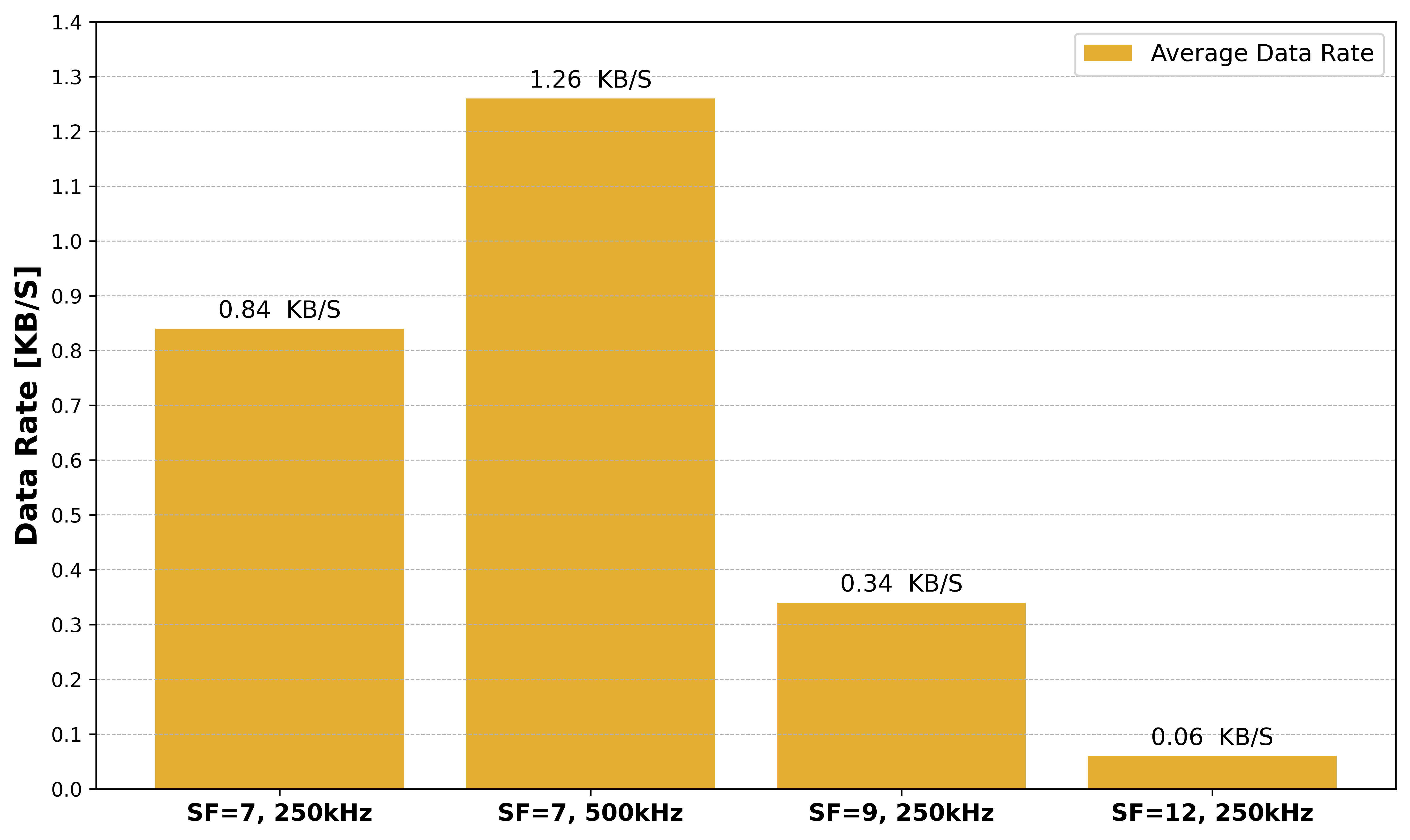} 
\caption{{\color{black}Data rates for various configurations of SF and BW at 100\% duty cycle}}
\label{fig: data_rate}
\end{figure}

\subsubsection{Response Time}

We performed a set of experiments and recorded the response time of the LoRaWeb system. Fig. \ref{fig: response_time_multi} depicts the response times achieved by the LoRaWeb system for different combinations of SF, BW, and duty cycle. We tested the system with various combinations of SF, BW, and duty cycle. The lowest response time of 0.118 seconds was observed for SF = 7 and a bandwidth of $500KH_z$ for all duty cycle settings. When the spreading factor was kept the same at 7, and the bandwidth was reduced to $250 KH_z$, the response time was increased to 0.217 S. This shows the impact of bandwidth on the response time of the LoRaWeb system. The response times increased with higher spreading factors as with SFs 9 and 12 with bandwidth set to $250KH_z$; the response times recorded were 0.647 seconds and 3.901 seconds, respectively. Duty cycles showed negligible impact on the response times and thus, it achieved constant response times for all duty cycle settings. 

\begin{figure}[h]
    \captionsetup{labelfont={color=black}}
    \centering
    \includegraphics[width=0.7\columnwidth]{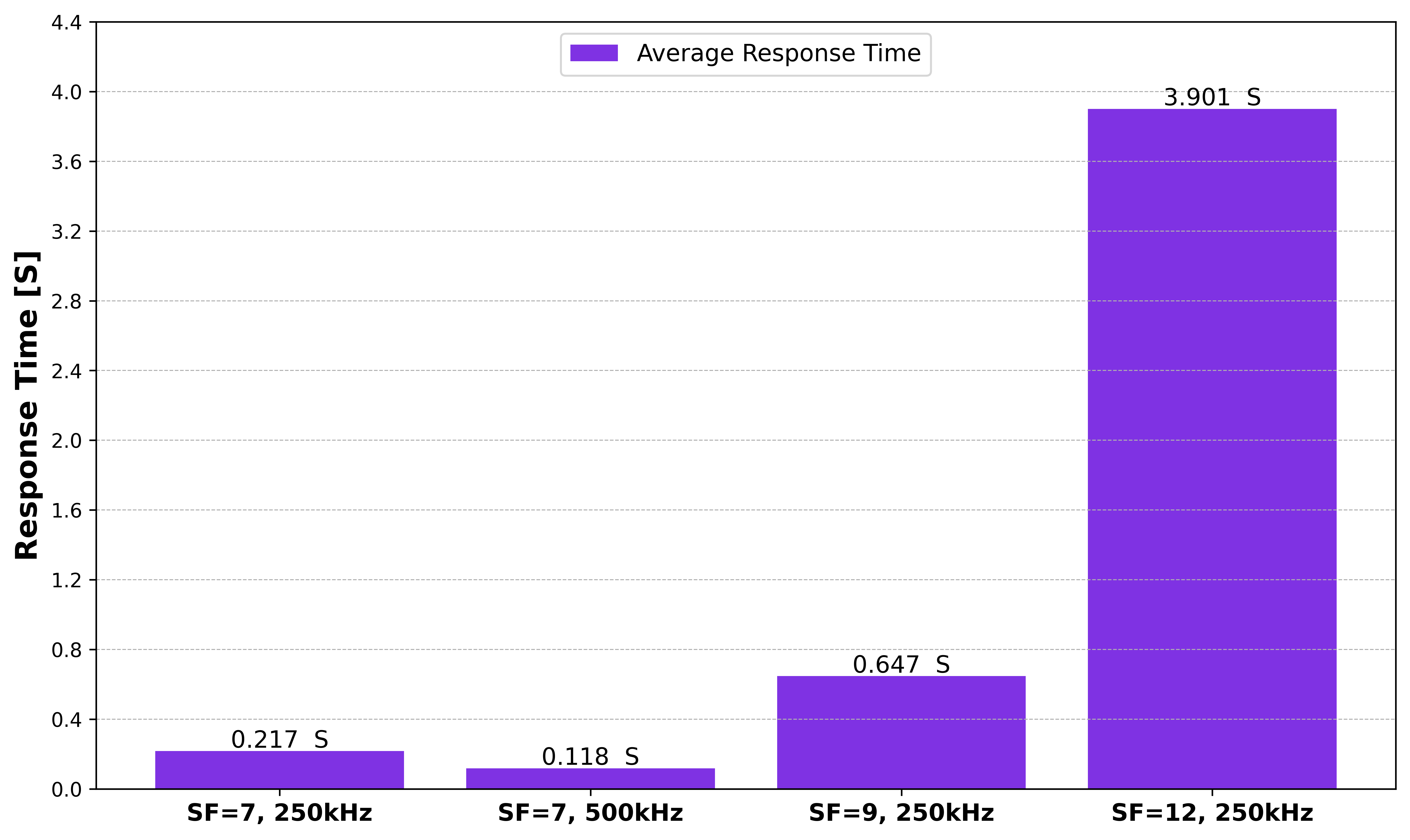}
    \caption{{\color{black}Response time for various configurations}}
    \label{fig: response_time_multi}
\end{figure}

\subsubsection{Access Delay}
We performed several rounds of experiments to assess the impact of bandwidth, spreading factor, and duty cycle on the web page access delay on the LoRaWeb system. The observations of our experiments are depicted in Fig. \ref{fig: access_time_multi}. The spreading factor and the corresponding bandwidth settings directly impacted the access delay. When a spreading factor of 7 and a bandwidth of $500KH_z$ were set, the LoRaWeb system achieved the lowest access delay. However, for the same spreading factor, as we reduced the bandwidth to $250 KH_z$, the access delay increased to 1.91 seconds. When we set higher spreading factors (9 and 12) on the LoRaWeb system, we observed a further increase in the access delay. The access delay increased to as high as 25.68 seconds when the spreading factor was set to 12.

\begin{figure}[h]
    \captionsetup{labelfont={color=black}}
    \centering
    \includegraphics[width=0.7\columnwidth]{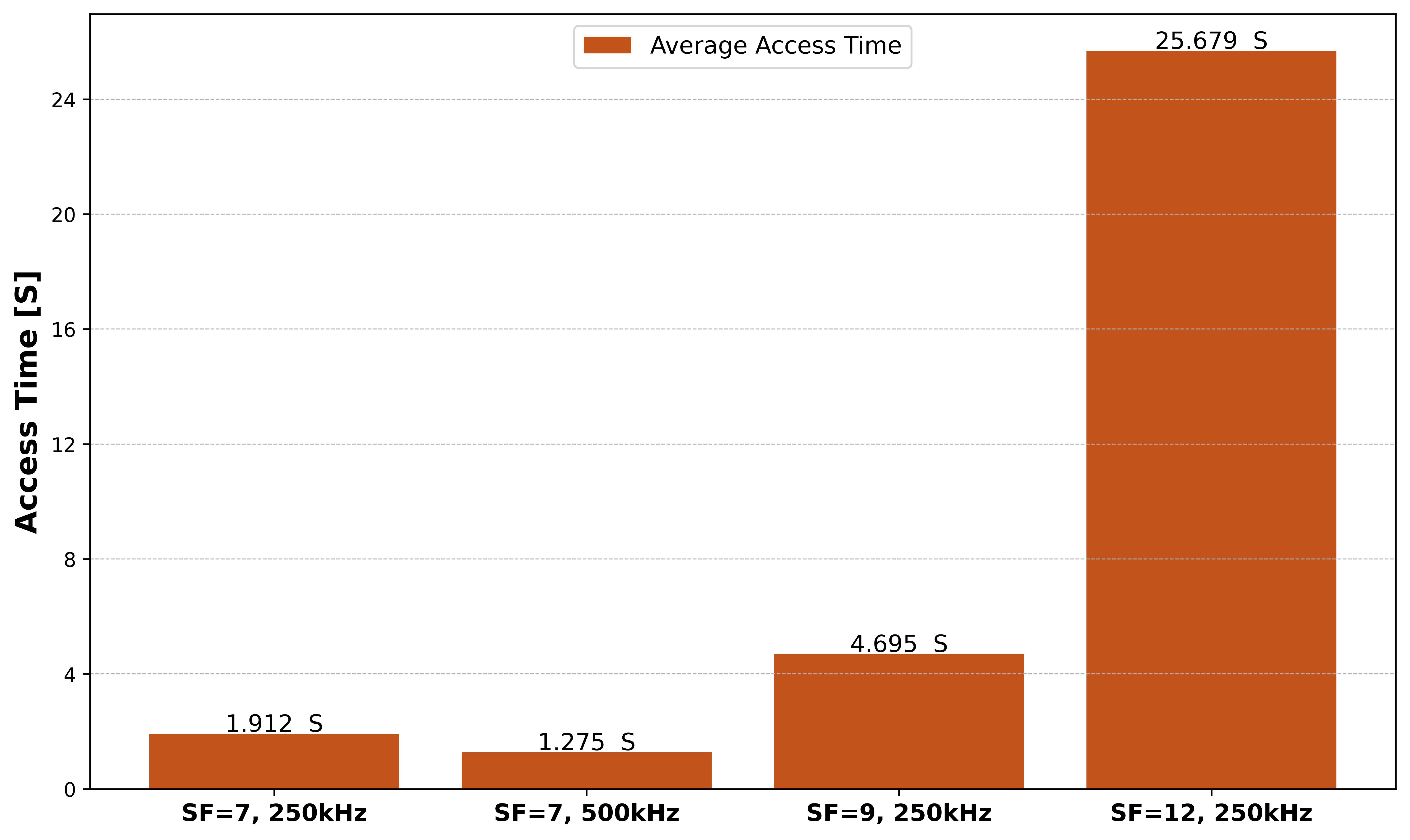}
    \caption{{\color{black}Access delay for various configurations of spreading factor, bandwidth, and duty cycle}}
    \label{fig: access_time_multi}
\end{figure}

\subsubsection{Environmental Impact}
The experimental results reveal the impact of environmental factors such as walls and electrical interferences on the performance of the LoRaWeb system. The LoRaWeb server was fixed on a pole in the experiment, as depicted in Fig. \ref{fig: system implementation}. Two client nodes were placed close to the server, and the other two client nodes were placed in different rooms on the same floor of the building. The walls acted as barriers to signal propagation. 

Signal attenuation and performance degradation were observed due to fading caused by multipath propagation and barriers like the walls of the rooms. RSSI values fluctuated between $-40 dBm$ to $-96 dBm$, and SNR values dropped below $-5 dB$ in high SF configurations. Reduced PDR such as $\approx 18.9\%$ at $SF = 12$ and $10\%$ duty cycle was also observed due to prolonged airtime of higher SFs, indicating susceptibility to fading.

Impact on transmission times was also observed, with lower SFs achieving shorter times than higher SFs. These results highlight the trade-off between long-range and environmental resilience.

\subsubsection{Scalability and Collision Impact}

We varied the configurations of the LoRaWeb system to assess its scalability and impact of collision. We observed that the system maintained the data rates and response times for each configuration of spreading factor, bandwidth, and duty cycle. For instance, it maintained a throughput of 1.18KB/S for all duty cycles when a spreading factor of 7 and a bandwidth of $500KH_z$ were used. This ascertains that the LoRaWeb system was able to handle high data transmission with minimal impact on the performance. Higher spreading factors on the other hand impacted the performance significantly. As a result, advanced strategies like FHSS and dynamic SF management are essential in dense networks.

\subsubsection{Cache Write and Read Time}\label{cache_fetch}

The LoRaWeb client node was set to write in the cache stored in the ESP32 file system. The experiment was repeated 20 times for data sizes of 1.5 KB, 3 KB, 4.5 KB, 6 KB, 7.5 KB, and 10 KB. The average times to write to the cache, as depicted in Fig. \ref{fig: cache_write_time}, were approximately 37 ms, 74 ms, 123 ms, 155 ms, 187 ms, and 243 ms, respectively. The cache write operation included opening a file in the ESP32 file system, writing to it, and closing it.  Additionally, the initial cache writes included the initialization of the cache system. Due to this, we observed a few cache write times to be higher than the rest. 

\begin{figure}[h]
\captionsetup{labelfont={color=black}}
\centering
\includegraphics[width=0.7\columnwidth]{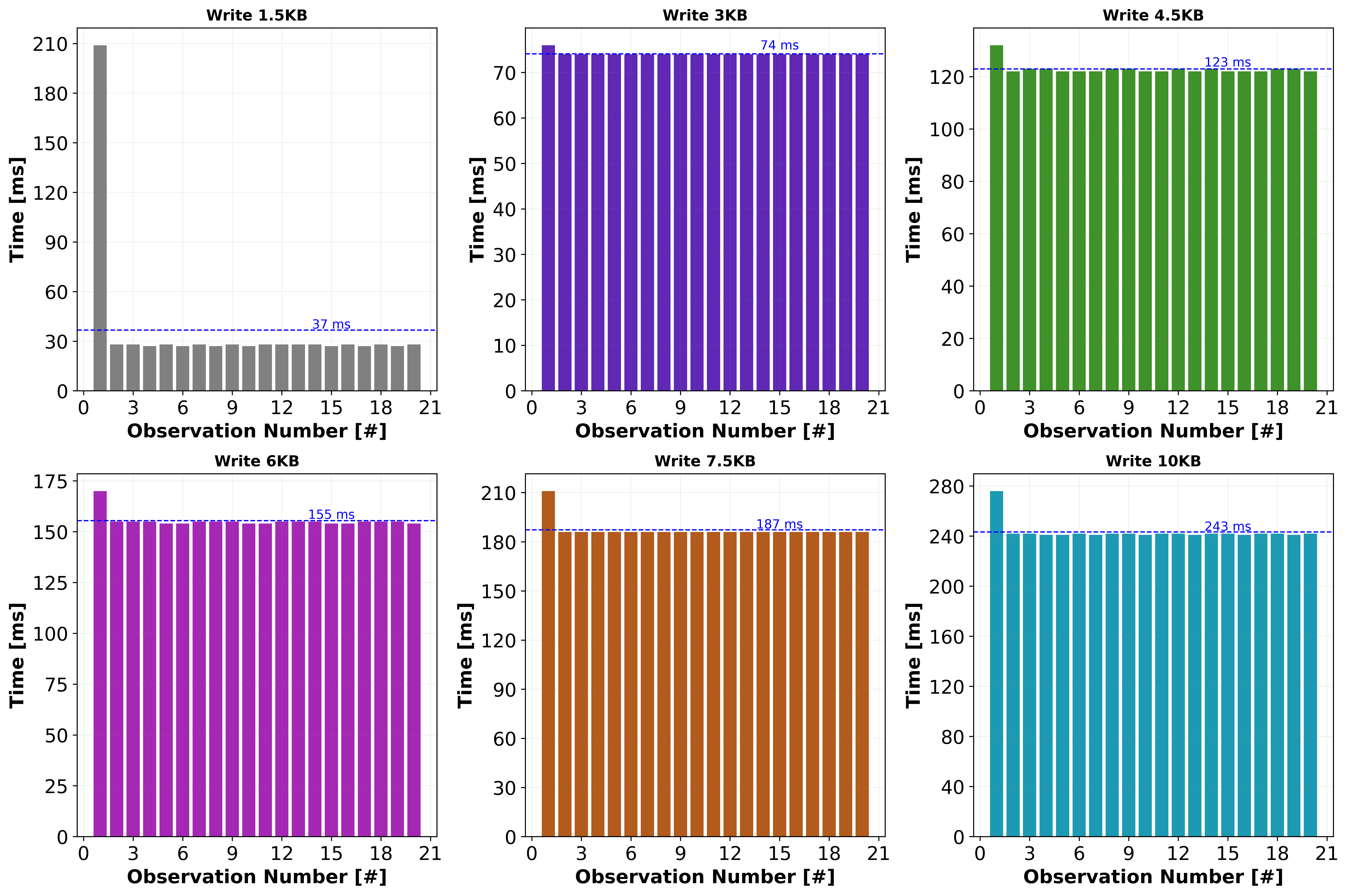}
\caption{{\color{black}Cache write time for varying sizes of data}}
\label{fig: cache_write_time}
\end{figure}

As in Fig. \ref{fig: cache_read_time}, we observed the cache read times to be lower than the write times, which were approximately 58 ms, 78 ms, 93 ms, 114 ms, 134 ms, and 187 ms, respectively. The low cache read times indicate the improvements in web page access delay and response times.

\begin{figure}[h]
\captionsetup{labelfont={color=black}}
\centering
\includegraphics[width=0.7\columnwidth]{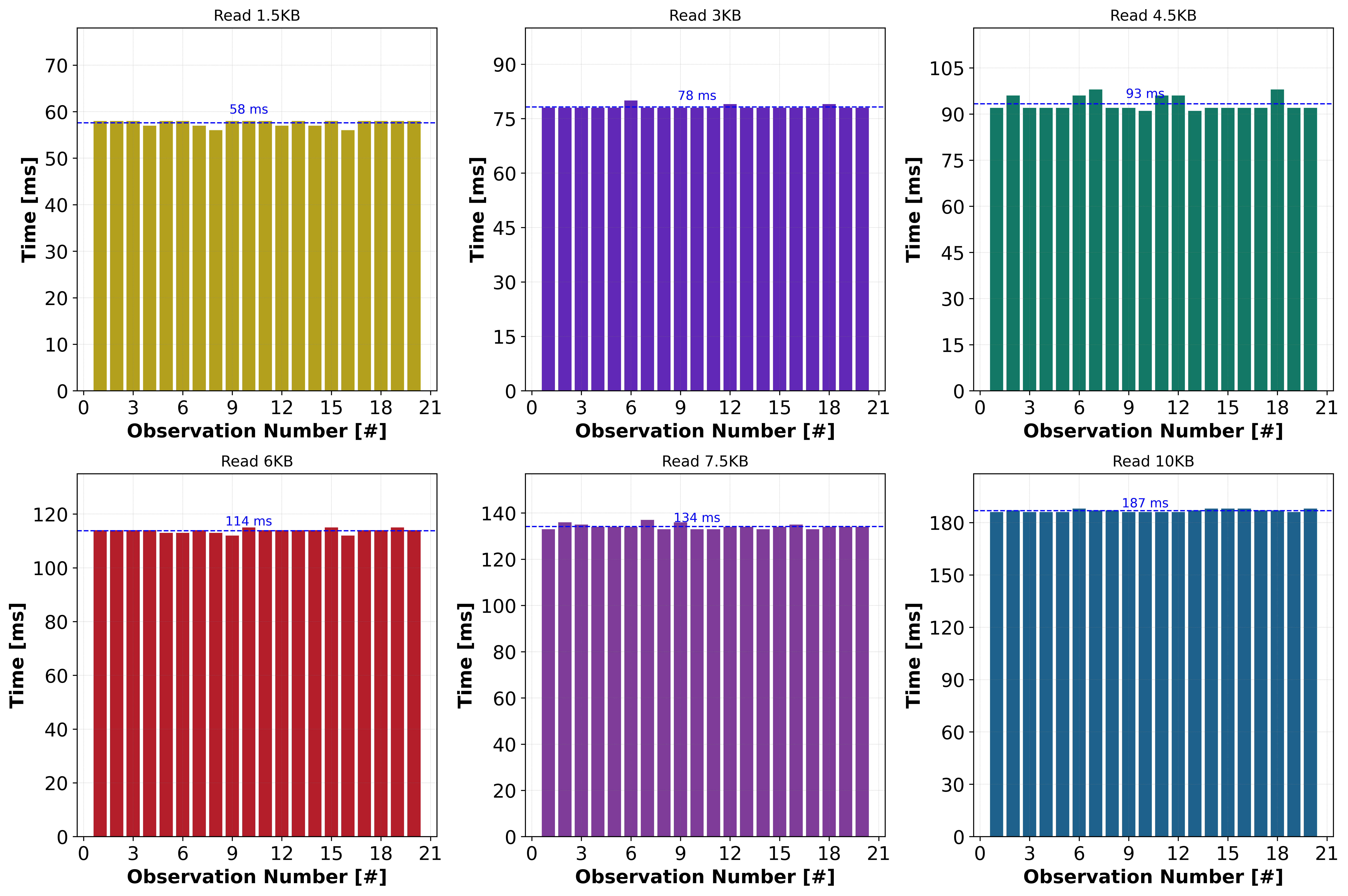} 
\caption{{\color{black}Cache read time for varying sizes of we pages}}
\label{fig: cache_read_time}
\end{figure}

\subsubsection{Cache Index Read and Write Time}\label{cache_index_lookup}
 
We evaluated the performance of the caching mechanism in the LoRaWeb system by measuring the times required to write and read the cache index. In our experiment, we evaluated the cache index write operation, which involved opening a file on the ESP32 file system, saving a small dictionary containing approximately $10 B$ of data, and then closing the file. The LoRaWeb system took an average time of 6.7 ms to write the cache index. The read operation, on the other hand, required opening the index file, loading the dictionary from it, and searching for the URL requested by the user. The average time for the read operation was approximately 12.45 ms as in Fig. \ref{fig: index_read_write_time}.  

\begin{figure}[h] 
\captionsetup{labelfont={color=black}}
\centering
\includegraphics[width=0.7\columnwidth]{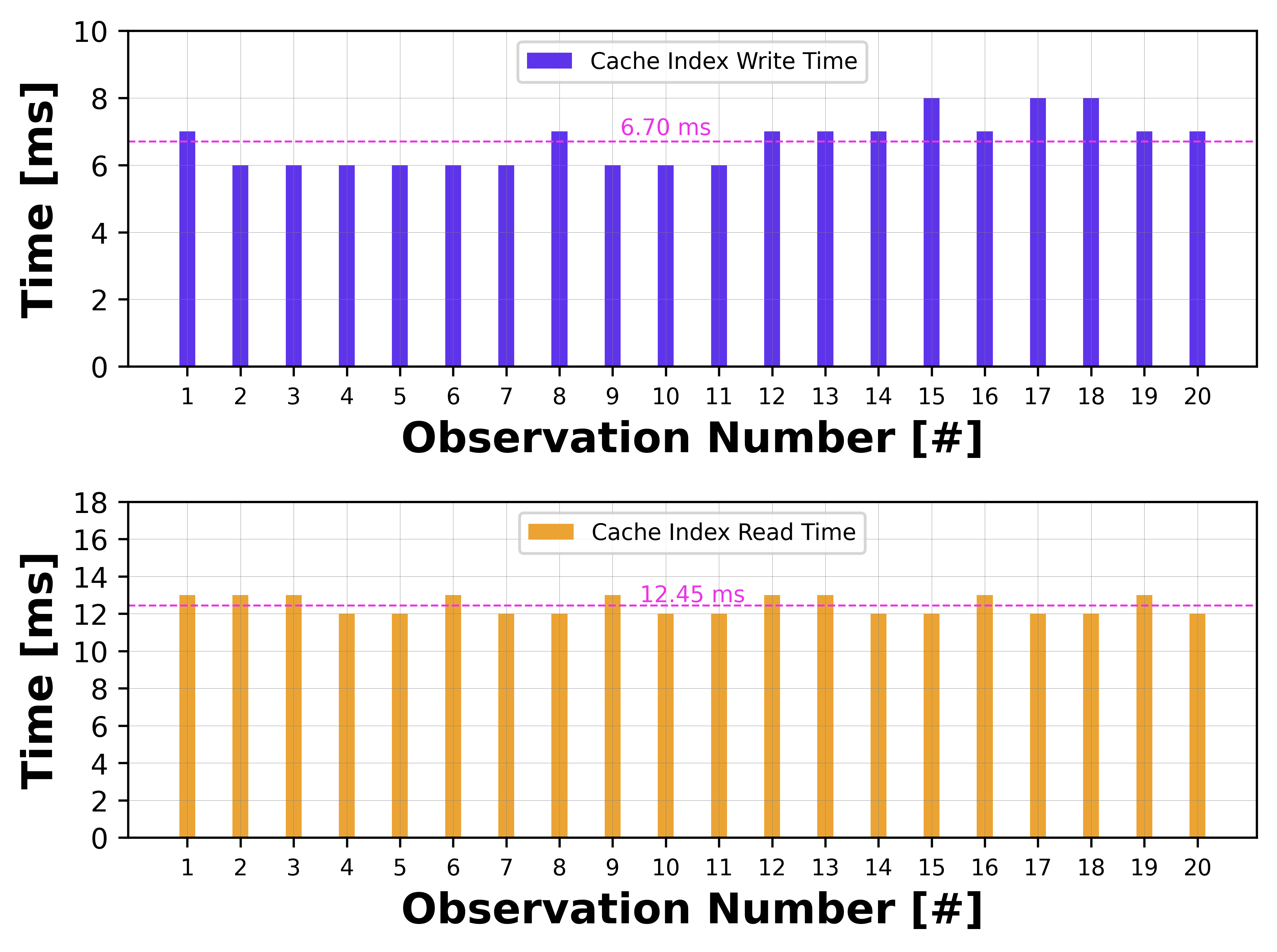}
\caption{{\color{black}{Cache index read and write times in ESP32 file system}}}
\label{fig: index_read_write_time}
\end{figure}

\subsubsection{Current Consumption}\label{current_consumption}

We measured the current consumed by the LoRaWeb system to assess its performance. For this, we set the client and the server nodes to exchange data to exchange 250 B sized messages. A USB tester (MX19) was plugged between the power source and the LoRaWeb nodes to measure the current consumption values. Figure \ref{fig: current_consumption} depicts the currents consumed by the server and the client node. The client node consumed higher current due to the presence of the additional WiFi radio in it. The LoRaWeb server and the client nodes consumed an average of 0.487 A and 0.551 A current, respectively.  

\begin{figure}[h] 
\captionsetup{labelfont={color=black}}
\centering
\includegraphics[width=0.7\columnwidth]{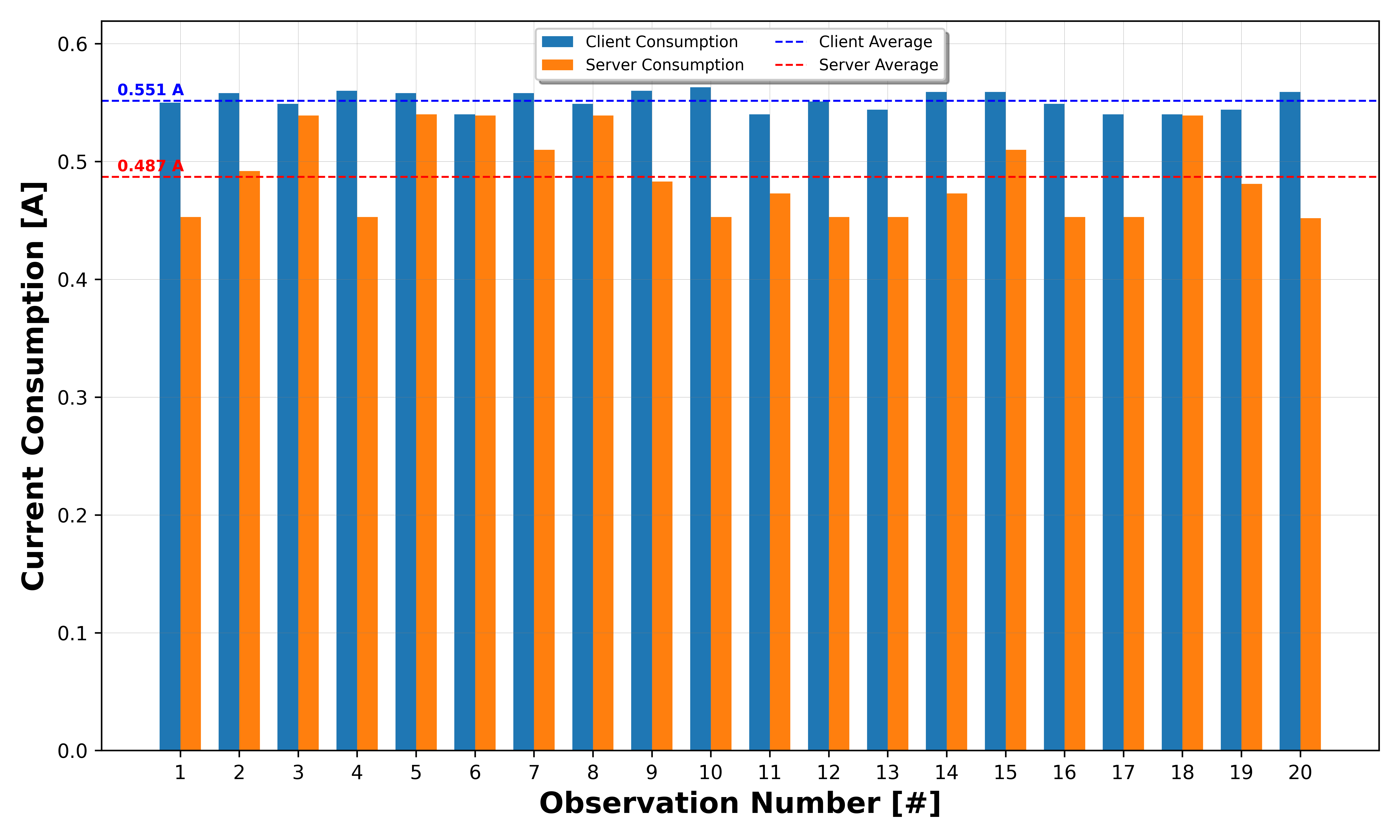}
\caption{{\color{black}{LoRaWeb node current consumption for data transmission and reception}}}
\label{fig: current_consumption}
\end{figure}

\subsubsection{Ping and Round Trip Time}\label{ping}

We implemented a lightweight ping functionality to test the extensibility and Internet potential of the LoRaWeb system. The average ping time was 174 ms and it varied between 68.82 ms and 284.33 ms, approximately.  Fig. \ref{fig: ping_rtt} depicts the recorded Round Trip Time (RTT) and Ping times in our experiment. The average RTT recorded was approximately 260 ms while it varied approximately between 160 ms and 370 ms. The results highlight the inherent limitations of LoRa affecting the performance. Accessing heavy web applications require additional exploration of innovative techniques. However, lightweight Internet communication over LoRa with satisfactory responsiveness is possible.

\begin{figure}[h]
    \captionsetup{labelfont={color=black}}
    \centering
    \includegraphics[width=0.7\columnwidth]{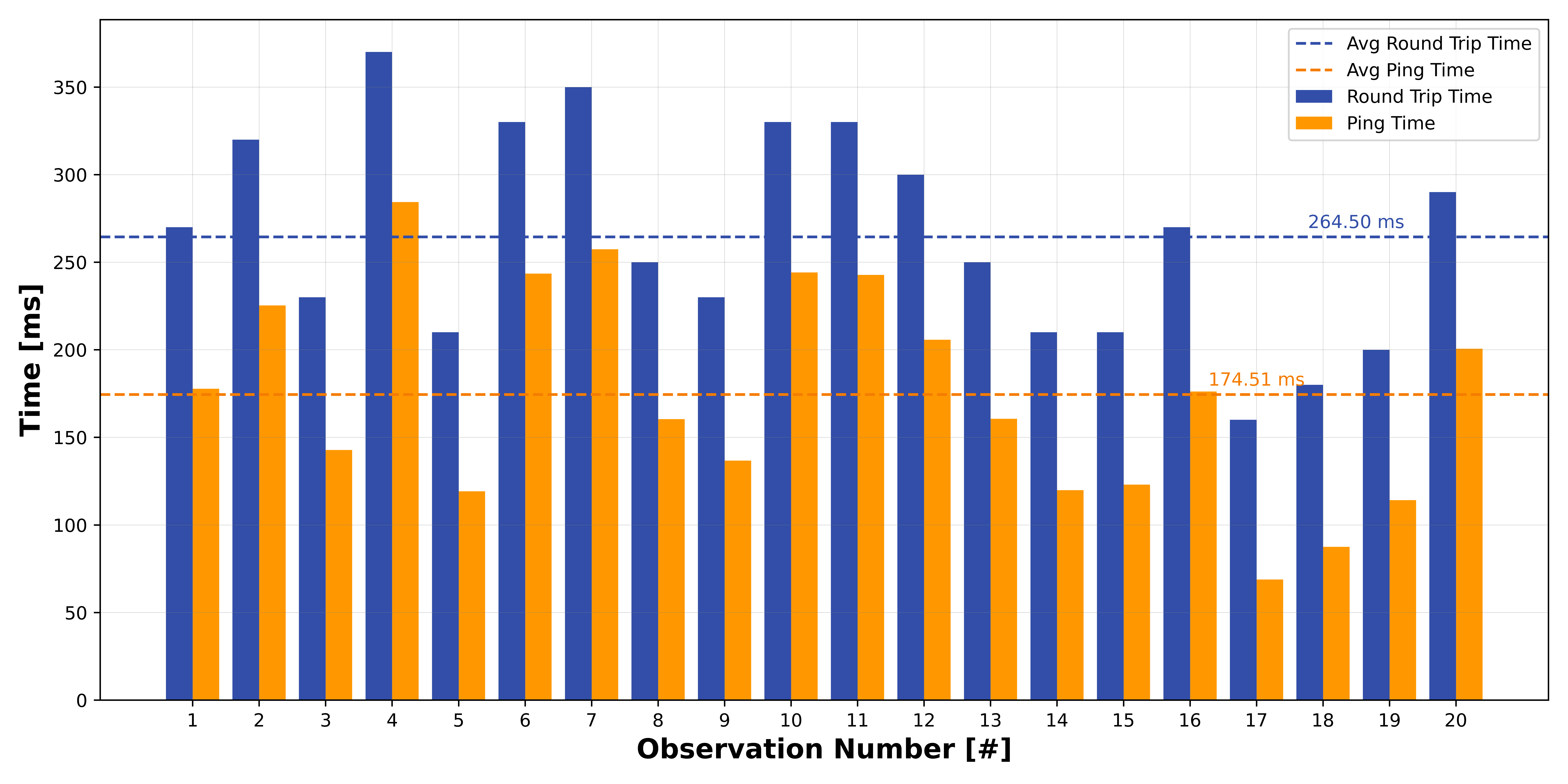}
    \caption{\color{black}{Ping and round trip times in LoRaWeb}}
    \label{fig: ping_rtt}
\end{figure}

\subsection{Setup-II: Single WiFi Client With A LoRaWeb Client}\label{single_wifi_client_setup}

This setup was configured to use $SF = 7$ and $BW = 500 kH_z$

\subsubsection{Web Page Access Delay} \label{request_response_delay}

The web page access delays were recorded while the experiment was conducted in $20$ rounds each for payload sizes of $1.5 KB$, $3 KB$, $4.5 KB$, $6 KB$, $7.5 KB$, and $10 KB$. The average access times were approximately $949 ms$, $1930 ms$, $2707 ms$, $3625 ms$, $4512 ms$, and $6739 ms$ respectively. The observations are depicted in Fig. \ref{fig: web_access_delay}. In contrast, the maximum access delays for the payloads were recorded as approximately $950 ms$, $1931 ms$, $2708 ms$, $3625 ms$, $4513 ms$, and $6740 ms$, respectively.

\begin{figure}[h]
    \captionsetup{labelfont={color=black}}
    \centering
    \includegraphics[width=0.6\columnwidth]{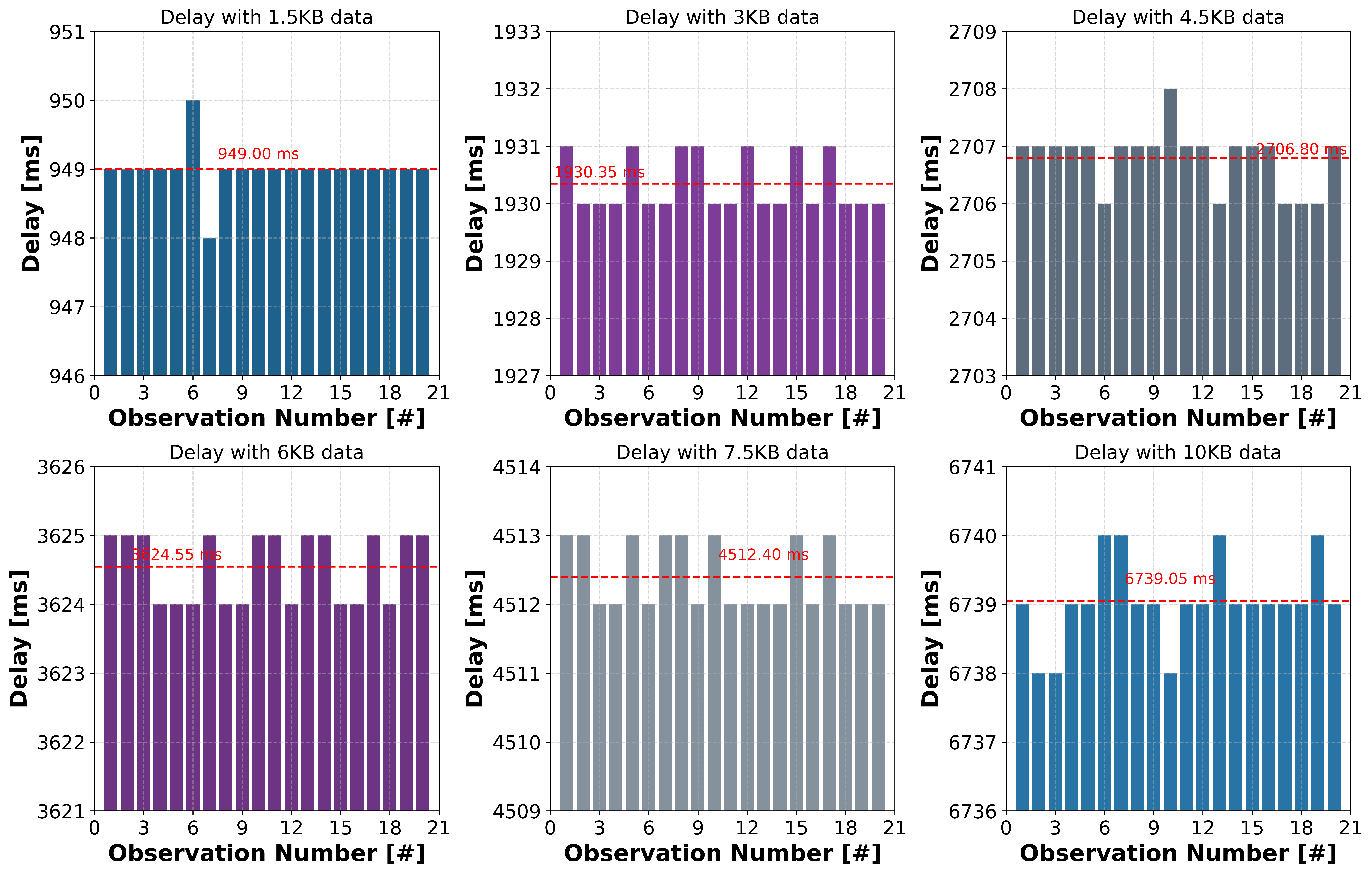}
    \caption{{\color{black}Access delay including re-transmission and synchronization delays}}
    \label{fig: web_access_delay}
\end{figure}

\subsubsection{Packet Loss} \label{packet_loss}
The efficiency of the proposed synchronization mechanism and acknowledgment handling was evaluated by conducting two rounds of experiments. $200$ messages were exchanged once with packet re-transmission in place and the other time without packet re-transmission. Figure \ref{fig: packet_loss} presents the detailed observations of the experiment. With re-transmission, all the packets were successfully transferred, and only $1$ was re-transmitted. In contrast, $197$ packets were successfully transmitted, and only $3$ packets were lost when there was no re-transmission. The synchronization mechanism efficiently handled the packet transmissions and achieves a higher Packet Delivery Ratio (PDR ($\lambda$)) as in Eq. \ref{eqn: pdr_1}.

$\lambda$ is given by the ratio of the number of packets successfully delivered ($\alpha$) to the total number of packets ($\beta$). Hence, the PDR with re-transmission implemented ($\lambda_1$) is given by - 

\begin{equation} \label{eqn: pdr_1}
    \lambda_1 = \frac{\alpha_1}{\beta_1} \times 100 = \frac{200}{200} \times 100 = 100\%
\end{equation}

Whereas, the PDR without re-transmission implemented ($\lambda_2$) is given by -
\begin{equation} \label{eqn: pdr_2}
    \lambda_2 = \frac{\alpha_2}{\beta_2} \times 100 = \frac{197}{200} \times 100 = 98.5\%
\end{equation}

\begin{figure}[h]
    \captionsetup{labelfont={color=black}}
    \centering
    \includegraphics[width=0.7\columnwidth]{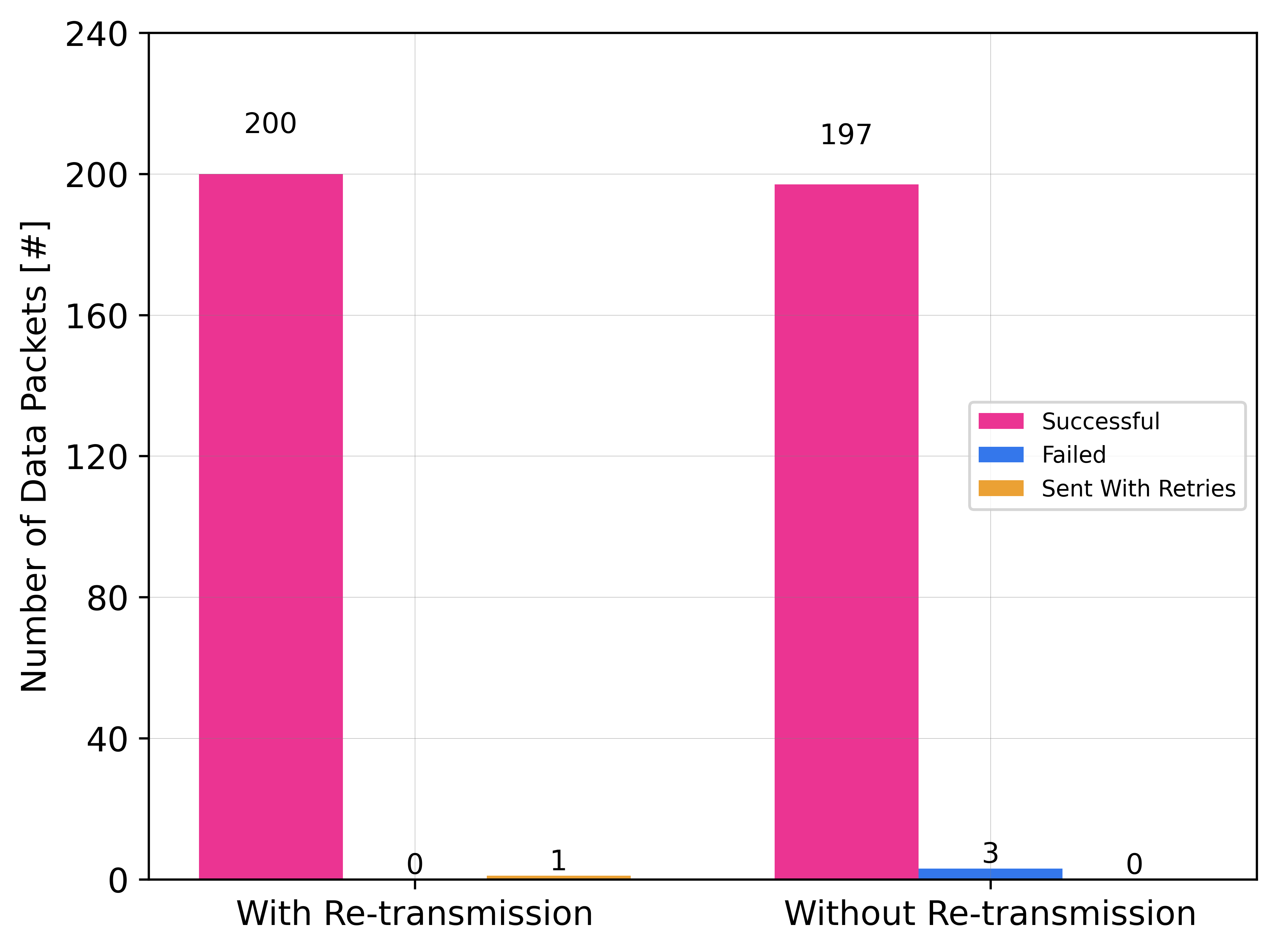}
    \caption{{\color{black}Packets successfully transmitted, lost, and re-transmitted}}
    \label{fig: packet_loss}
\end{figure}

\subsection{Setup-III: Multiple LoRaWeb Clients With A LoRaWeb Server Using FHSS}\label{fhss}

The static frequency communication was susceptible to interference and provided low reliability in high-traffic conditions. To address these limitations, we implemented and evaluated the FHSS with $SF = 7$ and $BW = 500 kH_z$. As in Algorithm \ref{fhss_client}, the system implemented a user-configurable number of retries, which was set to only three in this experiment. The experimental setup comprised a LoRaWeb server and four clients. Each client and the server were configured to operate over four predefined frequencies ($868.0 MH_z$, $868.2 MH_z$, $868.4 MH_z$, and $868.6 MH_z$) with a dedicated channel ($868.8 MH_z$) for acknowledgments. 

\begin{algorithm}
\textcolor{black}{
\caption{\textcolor{black}{FHSS Client Algorithm}}
\label{fhss_client}
\begin{algorithmic}[1]
    \renewcommand{\algorithmicrequire}{\textbf{Inputs:}}
    \renewcommand{\algorithmicensure}{\textbf{Output:}}
\Require LoRa module initialized, frequency list, client ID
\Ensure Successful data reception with frequency hopping
\State \texttt{seqNum} $\gets 0$
\While{\texttt{successCount} $<$ \texttt{MAX\_REQ}}
    \State Wait \texttt{random(MIN\_DLY, MAX\_DLY)}
    \State \texttt{msg} $\gets$ \texttt{"req \%s \%d"} (\texttt{ID}, \texttt{seqNum})
    \State \texttt{ack} $\gets$ \texttt{false}, \texttt{retry} $\gets 0$
    \While{\texttt{!ack} \textbf{and} \texttt{retry} $<$ \texttt{MAX\_RETRY}}
        \State Send \texttt{msg} on \texttt{freq}
        \State Switch to \texttt{ACK\_CH}, wait for \texttt{ACK}
        \If{\texttt{ACK received}}
            \State \texttt{ack} $\gets$ \texttt{true}
        \Else
            \State \texttt{retry++}
        \EndIf
    \EndWhile
    \If{\texttt{ack}} 
        \State Receive data while hopping channels
    \EndIf
    \State Hop to the next frequency
\EndWhile
\end{algorithmic}
}
\end{algorithm}

The system was configured to switch frequencies automatically. As a result, it achieved high reliability with $48$ successful requests out of $52$ attempts and only four retries. Thus significantly reducing collisions and the effect of interference and outperforming the static frequency setup (Section \ref{muli_loraweb_client}). The results highlight excellent improvements in packet delivery, data rates, and retry and suggest FHSS as a practical enhancement for IoT deployments that require low-power, reliable, and long-distance communication.


\section{Discussion and Limitations}
The results emphasize the need to carefully tune the SF and BW for a proper balance between range and data throughput while maintaining duty cycle limits. Configurations such as SF = 7 and BW = 500 kHz achieved an access delay of only $\approx 950ms$ for $1.5KB$ webpage, significantly outperforming Sigfox (2–3 seconds for a 12-byte payload) \cite{8660398} and the LoRa-based messaging system in \cite{10.1007/s11036-019-01235-5} taking $\approx 2.1S$ for unicast messages of much smaller payload. Furthermore, it offers greater infrastructure independence and scalability than NB-IoT, which relies on cellular networks. FHSS enhances the robustness and overall performance of the system.

For broader adoption in changing network conditions, fixed SF and BW settings remain a bottleneck. Extending the functionality to include complex functions such as web pages with multimedia transfer remains challenging. Furthermore, delivering public Internet-based applications in a longer range requires additional research.


\section{Conclusion} \label{conclusion}
We demonstrated that HTTP access is not only possible but also practical over resource-constrained LoRa networks. Methods were proposed to address challenges like low data rates, payload limitations, and multi-client contention with solutions such as message slicing, caching, and synchronization. We proved the effectiveness of the proposed system through actual hardware-based experiments. The experimental results showed that LoRaWeb delivers reliable performance and competitive response times, both in single-client and multi-client setups.
}

\bibliographystyle{IEEEtran}
\bibliography{references}

\end{document}